\documentclass{PoS}

\newcommand{\jpsi}{J/\psi}
\newcommand{\intq}{\int\!\frac{\mathrm d^3q}{(2\pi)^3}}
\newcommand{\mev}{~\mathrm{MeV}}
\newcommand{\gev}{~\mathrm{GeV}}
\newcommand{\jc}{{\cal J}}

\title{Properties of $\jpsi$ in light quark matter}

\ShortTitle{Properties of $\jpsi$ in light quark matter}

\author{\speaker{Martin Cleven}\\
        Departament de Fisica Quantica i Astrofisica and Institut de Ciencies del Cosmos\\
        Universitat de Barcelona, 08028-Barcelona, Spain\\
        E-mail: \email{cleven@fqa.ub.edu}}

\author{Volodymyr K. Magas\\
        Departament de Fisica Quantica i Astrofisica and Institut de Ciencies del Cosmos\\
        Universitat de Barcelona, 08028-Barcelona, Spain\\
       E-mail: \email{vladimir@fqa.ub.edu}}        
        
\author{Angels Ramos\\
        Departament de Fisica Quantica i Astrofisica and Institut de Ciencies del Cosmos\\
        Universitat de Barcelona, 08028-Barcelona, Spain\\
       E-mail: \email{ramos@fqa.ub.edu}}

\abstract{With various experiments studying heavy-ion collisions a demand exists in the hadron physics community for theoretical predictions of hadronic properties at temperatures and densities far from equilibrium. 
In this work we will study the $\jpsi$ vector meson at finite temperatures surrounded by light quark matter. 
We will apply a chiral unitary approach to account for coupled channels, most importantly channels with open charm. 
The in-medium solution accounts for the change in self-energy that the $\jpsi$ acquires from interacting with the surrounding light quark matter, most notably pions and rho mesons. 
The results are preliminary and clearly show the importance of using dressed charmed mesons. 
Ultimately, the solutions to the corresponding Lippmann-Schwinger Equations are used to calculate observables such as the spectral function of the $\jpsi$.}

\FullConference{VIII International Workshop On Charm Physics\\
		5-9 September, 2016\\
		Bologna, Italy}

\begin{document}

\section{Introduction}
%
With a new generation of heavy-ion-collision experiments -- LHC, FAIR or NICA -- operational or under construction there is a demand for theoretical predictions of hadronic properties at temperatures and densities far from equilibrium.
Among the various subjects the study of the $\jpsi$ stands out as a candidate to signal deconfinement. 
A quark-gluon plasma (QGP) produced in the collision would screen the $c\bar c$ interaction or ionize the charmonium state. 
Either way, this would lead to a suppression of events.
However, although in previous experiments such a drop was actually seen, it remains unclear whether this is indeed related to the formation of a QGP. 
The inelastic interactions of the $\jpsi$ with the surrounding medium, accounted for in the  'co-mover' models, offer another mechanisms to explain the drop in the $\jpsi$ production. Therefore, the modifications induced by a hot light quark matter on these interactions should be addressed to properly understand the transition to a QGP.


Previous studies on this subject broadly fall into two categories: interactions based on chiral Lagrangians~\cite{Haglin:2000ar,Bourque:2008ta,Blaschke:2008mu,Blaschke:2012zza} and quark model calculations~\cite{Zhou:2012vv,Maiani:2004py,Maiani:2004qj,Bourque:2008es}. Zhao and Rapp~\cite{Zhao:2010nk} use thermal lattice QCD to constrain the in-medium charmonium properties in order to calculate spectral functions of the $\jpsi$.
While those works have gone a long way in explaining the reduction of $\jpsi$ production in previous experiments, the calculations typically lack some of the state-of-the-art techniques that have been developed in the subsequent years. 
With this in mind and the need for updated predictions due to the next generation of experiments we will improve the calculations compared to previous studies in two major ways.
Firstly, we will use unitarized coupled channel amplidudes. This has a sizable impact on the dissociation cross sections for the  $\jpsi$. 
Secondly, we will use the Imaginary Time Formalism (ITF) to rigorously introduce temperature and many-body effects, while other studies often accounted for these with ad-hoc factors. 

This work is structured as follows: In Sec.~\ref{sec:su4} we will give an overview of the $SU(4)$ model developed by Gamermann et al.~\cite{Gamermann:2006nm,Gamermann:2007fi}, in Sec.~\ref{sec:itf} we will use IFT to extrapolate the results to finite temperatures and densities. We will finish with a short outlook and a brief summary. 

\section{Unitarized amplitudes in the vacuum}\label{sec:su4}
We will begin by laying out the foundations to calculate pseudoscalar-vector meson scattering in a chiral $SU(4)$ model. 
This model was succesfully applied in various works -- see $e.g.$ Refs.~\cite{Gamermann:2006nm,Gamermann:2007fi}. 
We will only focus on the essential points here and refer to these references for further reading. 
The pseudoscalar and vector fields are collected in the $SU(4)$ 15-plets $\Phi$ and $\mathcal V_\mu$, respectively, which are defined as

{\begin{footnotesize}
\begin{eqnarray}
\Phi= \left( \begin{array}{cccc}
\frac{1}{\sqrt{2}}\pi^0+\frac{1}{\sqrt{3}}\eta & \pi^+ & K^+ & \bar D ^0\\
\pi^- & -\frac{1}{\sqrt{2}}\pi^0+\frac{1}{\sqrt{3}}\eta &  K^0 & D^-\\
K^- & \bar{K^0} & \sqrt{\frac{2}{3}}\eta  & D_s^- \\
D^0 & D^+ & D_s^+ & \eta_c
\end{array} \right), \quad
\mathcal V_\mu= \left( \begin{array}{cccc}
\frac{1}{\sqrt{2}}\rho^0_\mu+\frac{1}{\sqrt{2}}\omega_\mu & \rho^+_\mu & K_\mu^{*+} & \bar D_\mu ^{*0}\\
\rho^-_\mu & -\frac{1}{\sqrt{2}}\rho_\mu^0+\frac{1}{\sqrt{2}}\omega_\mu &  K_\mu^{*0} & D_\mu^{*-}\\
K^{*-} & \bar{K}_\mu^{0*} & \phi_\mu & D_{s\mu}^{*-} \\
D_\mu^{*0} & D_\mu^{*+} & D_{s\mu}^{*+} & \psi_\mu\end{array} \right).
\end{eqnarray}
\end{footnotesize}
}

For each of the two one can construct a vector current
\begin{eqnarray}
  J_\mu = (\partial_\mu\Phi)\Phi - \Phi(\partial_\mu\Phi) ,\qquad
  \jc_\mu = (\partial_\mu \mathcal V_\nu)\mathcal V^\nu - \mathcal V_\nu(\partial_\mu \mathcal V^\nu). 
\end{eqnarray}
Connecting the two we can obtain the interaction Lagrangian:
\begin{equation}
 \mathcal{L} = -\frac{1}{4f^2}\left<J^{\mu}  \jc_\mu \right>
\end{equation}
where $<..>$ denotes the trace over flavor indices and $f=93\mev$ is the pion decay constant. 
Since $SU(4)$ is not an exact symmetry in nature one needs to break the symmetry in an appropriate way. 
This is done with factors that correspond to the leading allowed $t$-channel exchange -- $\gamma = (m_L/m_H)^2$ for charmed mesons and $\psi = -1/3+4/3(m_L/m'_H)^2$ for charmonia. 
We will use $m_L$=800 MeV, $m_H$=2050 MeV and $m'_H$=3~GeV in agreement with previous works. For details on this procedure see Refs.~\cite{Gamermann:2006nm,Gamermann:2007fi}.
The resulting amplidudes for pseudoscalar-vector scattering derived from this Lagrangian have the form
\begin{equation}\label{Eq:V}
 V_{ij}(s,t,u) = - \frac{\xi _{ij}}{4f^2}(s-u) \epsilon\cdot \epsilon', 
 \qquad
 \xi = \left( \begin{array}{ccc}  0 & 0 & \sqrt{8/3}\gamma \\  0 & 0 & \sqrt{8/3}\gamma \\  \sqrt{8/3}\gamma & \sqrt{8/3}\gamma & -\psi   \end{array}\right)
\end{equation}
where the indices $i,j$ cover the relevant meson pairs with quantum numbers $J^{PC}(I^G)=1^{++}(1^+)$ and hidden charm -- $J/\psi \pi$, $\eta_c\rho$ and $DD^*$. 

The corresponding $S$-wave projections will be used as the kernel for the Bethe-Salpeter equation. 
Using the on-shell formalism this simplifies to a simple algebraic equation that can easily be solved as
\begin{equation}
 T= (1 - VG)^{-1}V \vec \epsilon\cdot \vec \epsilon'
\end{equation}
\begin{figure}[t]
\centering
\begin{minipage}{0.49\linewidth}
 \includegraphics[width=\linewidth]{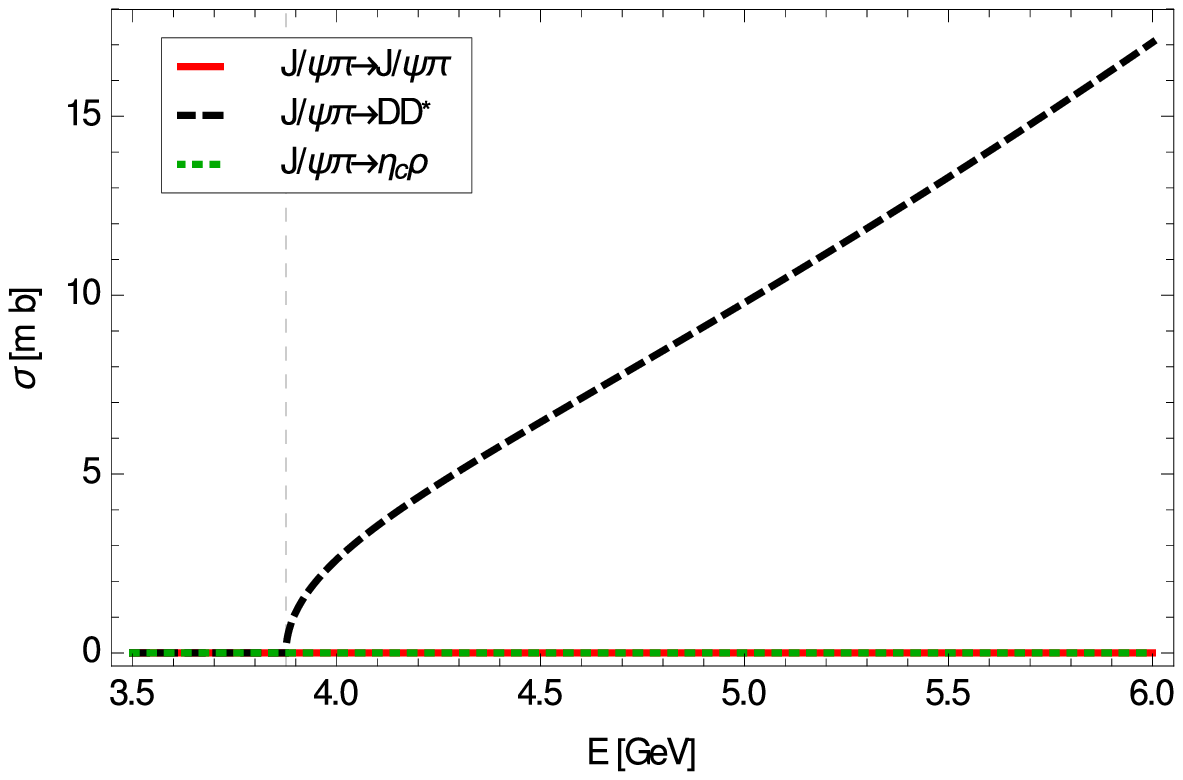}
\end{minipage}
\begin{minipage}{0.49\linewidth}
 \includegraphics[width=\linewidth]{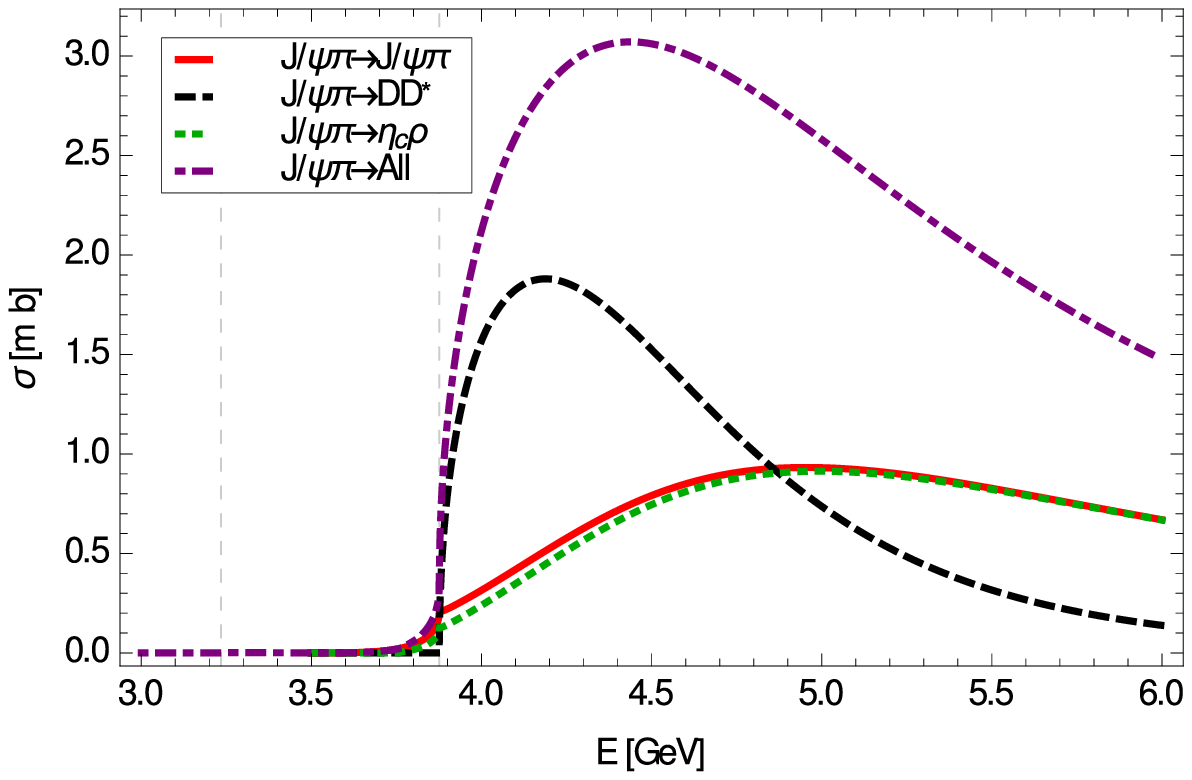}
\end{minipage}
\caption{Cross Sections for $J/\psi\pi\to X$ employing the non-unitarized amplitudes (left panel) and the unitarized ones (right panel)} \label{fig:JpsipiX_Vacuum}
\end{figure}
The diagonal matrix $G$ contains the two-meson loops which can be calculated analytically as
\begin{eqnarray}\label{Eq:G_Vacuum}
 G_{ii}(s) &=& i \int\!\frac{\mathrm d^4q}{(2\pi)^4} \frac{1}{[q^2-m_1^2+i\epsilon][(P-q)^2-m_2^2+i\epsilon]}
\end{eqnarray}
where the index $i$ refers to the pair of meson with masses $m\,_1$ and $m\,_2$ and the center of mass momentum is given by $P\,^2 = s$. 
The loop integrals are calculated using dimensional regularisation. 
We will fix the free parameters following Ref.~\cite{Gamermann:2007fi}, $e.g.$ the subtraction constant is $\alpha\,_H=-1.55$ at the scale $\mu=1.5\gev$ to reproduce the $D^*_{s0}(2317)$ in the open charm and open strangeness sector with quantum numbers  $J^{P}=1^{+}$ and $I=0$. 

Fig.~\ref{fig:JpsipiX_Vacuum} shows the effect of unitarization on the cross sections for the processes $J/\psi\pi \to J/\psi\pi$, $J/\psi\pi \to \eta_c\rho$ and $J/\psi\pi \to DD^*$.
The most notable aspect is obviously the lack of $J/\psi\pi \to J/\psi\pi$ scattering at leading order for the cross section using non-unitarized amplitudes (left panel). 
This is an immediate consequence of the interpretation of the pion as the Goldstone boson of QCD and is also reflected by the vanishing leading order amplitude in Eq.~(\ref{Eq:V}).
As a result, all the processes occur through the charmed meson pairs. This will also become an important feature once we consider in-medium effects as we will see in the following chapter. 
Although slightly different in shape, the size of our unitarized cross sections is in line with what has been found with the chiral phenomenological models developed in  \cite{Haglin:2000ar,Bourque:2008ta}, the extended Nambu-Jona-Lasinio model \cite{Bourque:2008es},  a meson-exchange model \cite{Oh:2000qr}, a non-relativistic quark potential model \cite{Wong:2001td}, or a QCD-sum-rule approach \cite{Duraes:2002px} .

\section{Interactions at finite temperatures}\label{sec:itf}
We will study the deviation from the vacuum in two ways. 
First, we will use the Imaginary Time Formalism (ITF) to study the impact of finite temperatures. 
Secondly, we will dress the $\jpsi$ with the pion induced self energy to account for the effects of the mesonic matter surrounding it. Both need to be incorporated in a self-consistent way. 

We will briefly review the ITF, the details of which can be seen $e.g.$ in Refs.~\cite{galekapustabook,lebellac}. 
Essentially, the modifications of the Feynman rules apply to the propagators only while the vertices remain unchanged. The necessary modifications are
\begin{equation}
 q_0\to i\omega_n = i2\pi nT, \qquad \int\!\frac{\mathrm d^4q}{(2\pi)^4}\to iT\sum_n \int\!\frac{\mathrm d^3q}{(2\pi)^3}
\end{equation}
with the discrete Matsubara frequencies $\omega_n = i2\pi nT$. Once the sums over the Matsubara frequencies are performed one may analytically continue the result to external frequencies $\omega+i\epsilon$.

We will start by calculating the meson-meson loop at finite temperature. Applying  the ITF Feynman rules we find 
\begin{equation}
 G_{MM'}(W_m,\vec p;T) = - T\sum_n \intq D_M(\omega_n,\vec q;T)D_{M'}(W_m-\omega_n,\vec p-\vec q;T).
\end{equation}
where the meson propagator in a hot medium is given by
\begin{equation}
 D_M(\omega_n,\vec q;T) = [(i\omega_n)^2-\vec q^2-m_M^2-\Pi_M(\omega_n,\vec q;T)]^{-1}
\end{equation}
To carry out the Matsubara sum it is convenient to use the spectral (Lehmann) representation
\begin{equation}
 D_M(\omega_n,\vec q;T) = \int\!\mathrm d\omega \frac{S_M(\omega,\vec q;T)}{i\omega_n-\omega}
\end{equation}
The Matsubara sum yields
\begin{equation}
T \sum_n \frac{1}{[i\omega_n-\omega][iW_m-i\omega_n-\Omega]} = - \frac{1+f(\omega,T)+f(\Omega,T)}{i W_m- \omega-\Omega}
\end{equation}
and we find
\begin{eqnarray}\label{Eq:G_TVac}
 G_{MM'}(W_m,\vec p;T) = \intq \int\!\mathrm d\omega\int\!\mathrm d\Omega \frac{S_M(\omega,\vec q;T)S_{M'}(\Omega,\vec p - \vec q;T)}{iW_m-\omega-\Omega}     [1+f(\omega,T)+f(\Omega,T)]
\end{eqnarray}
with $f(\omega,T)=[\exp(\omega/T)-1]^{-1}$ the meson Bose distribution function at temperature $T$. 
This result can be extrapolated to the real axis using $iW_m\to p^0+i\epsilon$. 
For vanishing self energies we can further simplify the expression as the spectral function becomes the delta distribution:
\begin{equation}
 S_M(\omega,\vec q;T)\to \frac{\omega_M}{\omega}\delta(\omega^2-\omega_M^2)
\end{equation}
with $\omega_M=\sqrt{q_M^2+m_M^2}$.

\begin{figure*}[t]
\centering
 \includegraphics[width=\linewidth]{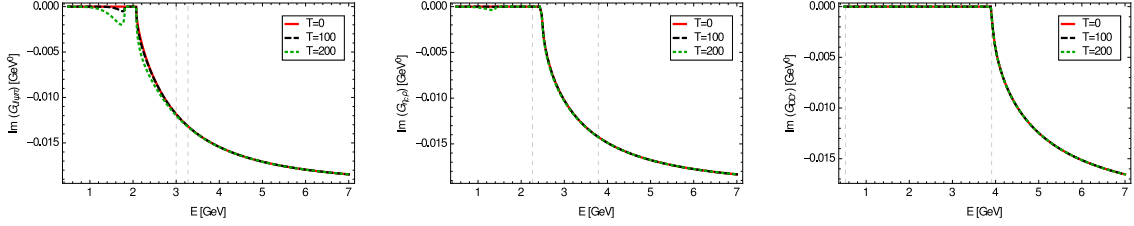}
\caption{Imaginary part of the two-body loops with the meson pairs $DD^*$, $\eta_c\rho$ and $J/\psi\pi$ at temperatures ${T=(0,100,200)\mev}$ } \label{fig:Loops_Temperature}
\end{figure*}

We will use a sharp cutoff to regularize the three-momentum integration here. 
However, to be consistent with Eq.~(\ref{Eq:G_Vacuum}) we will follow the approach by Tolos et al. in Ref.~\cite{Tolos:2009nn}. 
We will take the original calculation, performed in dimensional regularisation and without temperature dependence, $G(s)$ from Eq.~(\ref{Eq:G_Vacuum}), and correct it by a temperature-dependent function $\delta G(s,T)$ defined as
\begin{equation}\label{Eq:G_Correction}
 \delta G (s,T) = \lim_{\Lambda\to\infty} \delta G_\Lambda (s,T) = \lim_{\Lambda\to\infty} [ G_\Lambda (s,T) - G_\Lambda (s,T=0) ]
\end{equation}
Using $G(s,T)=G(s)+\delta G_\Lambda (s,T)$ allows us to smoothly continue from the original framework. 
At the same time, this drastically reduces the cutoff dependence of the results. 

In Fig.~\ref{fig:Loops_Temperature} we see the effect of increasing the temperature on the loops. 
As it turns out, the relative proximity of the pion mass to the temperatures used here ensures that the $\jpsi\pi$ loop sees significant impact from finite temperatures. 
At the same time nothing similar is found for the other two pairs, $\eta_c\rho$ and $DD^*$. There an effect becomes only visible when one goes to temperatures that are well beyond a physically meaningful scenario. 
Therefore, the resulting $T$-matrix has only a weak dependence on temperature, since it is mainly driven by the $DD^*$ loop because of the vanishing leading order amplitude for elastic $\jpsi\pi$ scattering (c.f. Eq.~(\ref{Eq:V})).
It will then be essential to study the effect that charmed meson propagators, dressed by the pion bath, have on the $DD^*$ loop. 
However, this goes beyond the scope of this work and will be subject of a subsequent, more comprehensive study. 
Within the current framework we find a $T$-matrix $T_{\jpsi\pi\to\jpsi\pi}$ that is essentially independent of the temperature. 

\begin{figure*}
\centering
\begin{minipage}{0.49\linewidth}
 \includegraphics[width=\linewidth]{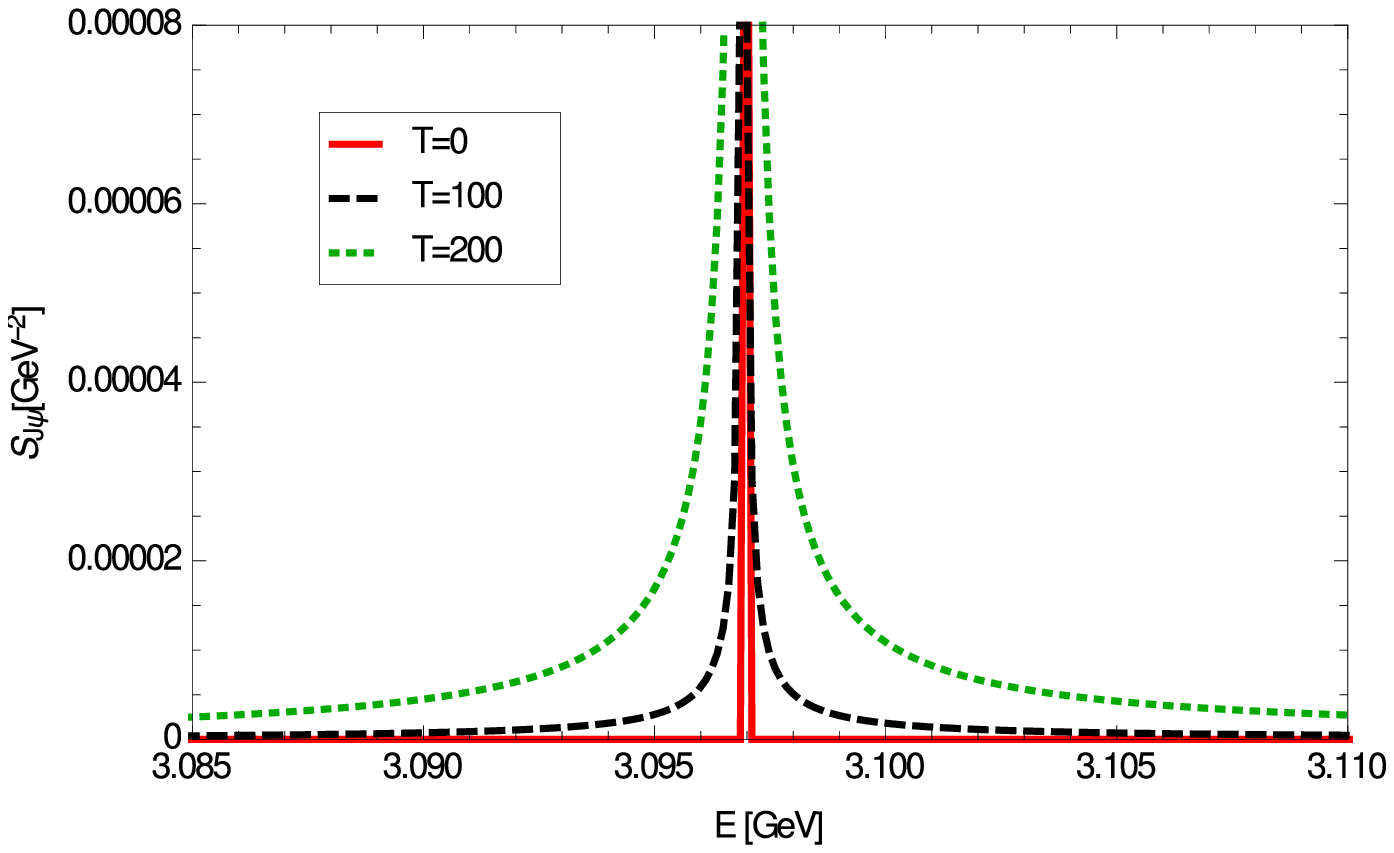}
\end{minipage}
\begin{minipage}{0.49\linewidth}
 \includegraphics[width=\linewidth]{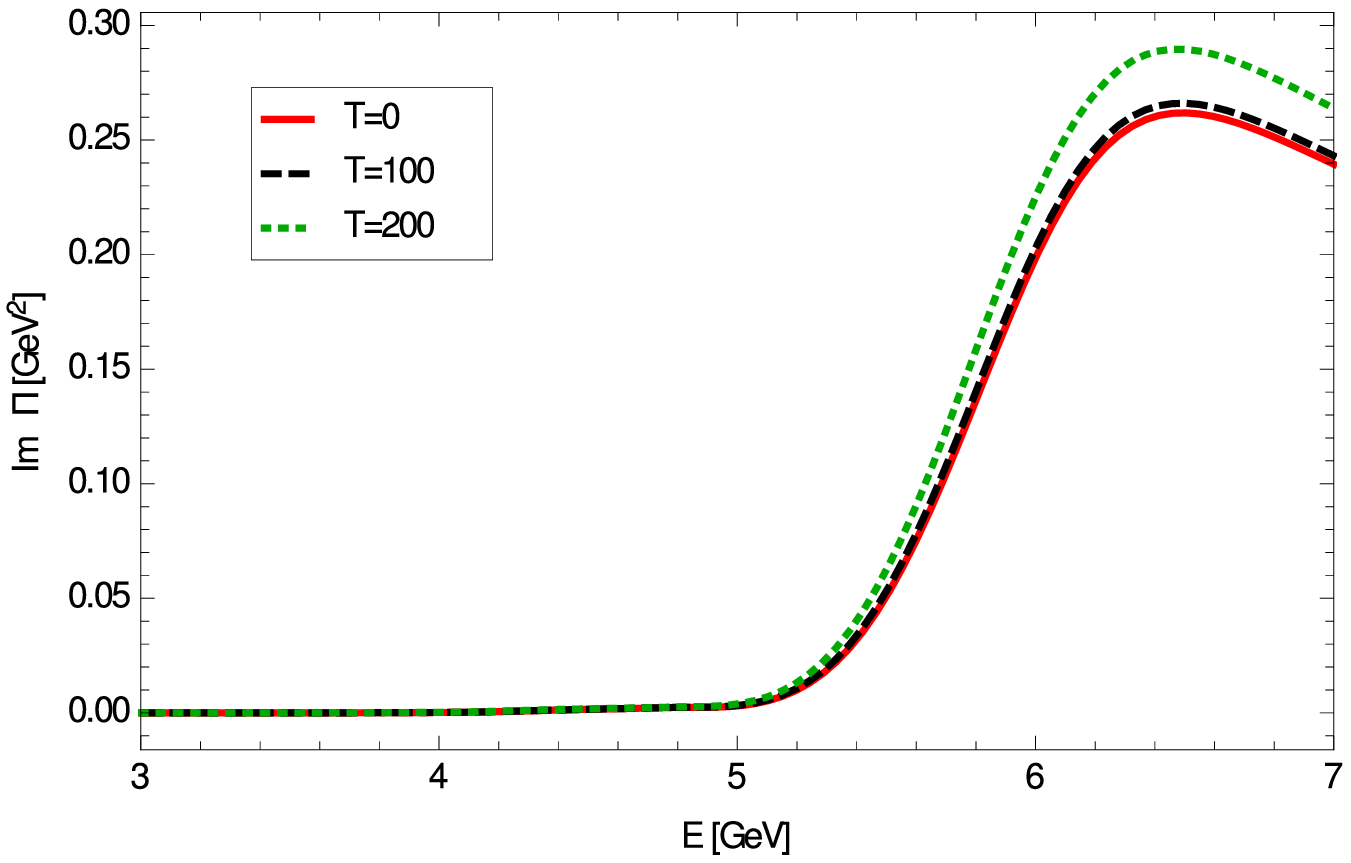}
\end{minipage}
\caption{
Left panel: Spectral function for the $\jpsi$ at rest, 
Right panel: Imaginary part of the $\jpsi$ self-energy from the pion loop. 
Both are shown for different temperatures $T=(0,100,200)\mev$
} \label{fig:SelfEnergy}
\end{figure*}
We can obtain the $J/\psi$ self energy from closing the pion line in the $T$-matrix for $\jpsi\pi\to\jpsi\pi$. Applying  the ITF Feynman rules we find
\begin{equation}
 \Pi_{J/\psi }(W_m,\vec p) = T\int\!\frac{\mathrm d^3q}{(2\pi)^3}\sum_n D_\pi(\omega_n,\vec q;T)T_{J/\psi \pi}(\omega_n+W_m,\vec p+\vec q;T)
\end{equation}
Rewriting this in terms of spectral representations for both the pion propagator and the $J/\psi \pi$ $T$-matrix, we
carry out the Matsubara sum and, using the corresponding delta distribution for the bare pion, we find for real energies $p^0$
\begin{eqnarray}\label{eq:Pi}
 \Pi_{J/\psi }(p^0,\vec p) = \int\!\frac{\mathrm d^3q}{(2\pi)^3} \int\!\mathrm d\Omega
 \frac{ f(\Omega,T)-f(\omega_\pi,T)}{(p^0)^2 - (\omega_\pi-\Omega)^2 + i\epsilon}  
 \left(-\frac1\pi\right) \mathrm{Im} T_{J/\psi \pi}(\Omega,\vec p+\vec q;T). 
\end{eqnarray}
In Fig.~\ref{fig:SelfEnergy} we show the results for self energy of the $\jpsi$ and the resulting spectral function.
Notice that the effect from the temperature stems almost solely from the  Bose distribution functions in Eq.~(\ref{eq:Pi}) since the temperature dependence of the $T$-matrix is negligible as was discussed before.
A larger effect at a physically relevant temperatures can be expected once the dressed charmed meson propagators are included. 

\section{Summary}
We have calculated the behaviour of $\jpsi$ in a pion bath at finite temperatures using unitarized $SU(4)$ chiral amplidudes. 
While there is already a visible effect from dressing the $\jpsi$ alone it is clear that a more comprehensive calculation involving dressed charmed mesons is necessary due to the fact that the amplidude for $\jpsi\pi\to\jpsi\pi$ is driven by the $DD^*$ loop. This will be the subject of a subsequent, more elaborate effort. 
Compared to previous studies on the subject this work is put on theoretically more sound foundations by using unitarized amplidudes and the Imaginary Time Formalism. 
The aforementioned use of unitarized amplitudes with coupled channels allows for the $\jpsi\pi$ pair to decay into $\jpsi\pi$ and $\eta_c\rho$. 
Both channels were not available to previous calculations. 
The resulting width can be used as input in theoretical simulations of $\jpsi$  propagation in a hot pion gas.

\end{document}